\documentclass[modern,trackchanges]{aastex61}

\shorttitle{ELECTROSTATIC WAVES IN RADIO ZEBRA MODELS} \shortauthors{Ben\'a\v{c}ek and Karlick\'y}
\begin{document}

\title{GROWTH RATES OF THE ELECTROSTATIC WAVES IN RADIO ZEBRA MODELS}

\author[0000-0002-4319-8083]{Jan Ben\'a\v{c}ek}
\affil{Department of Theoretical Physics and Astrophysics, Masaryk University,
   Kotl\'a\v{r}sk\'a 2, CZ-61137 Brno, Czech Republic}
\email{jbenacek@physics.muni.cz}

\author[0000-0002-3963-8701]{Marian Karlick\'y}
\affil{Astronomical Institute of the Academy of Sciences of the Czech Republic,
CZ-25165 Ond\v{r}ejov, Czech Republic}

\begin{abstract}
Zebras were observed not only in the solar radio emission but also in radio
emissions of Jupiter and Crab Nebula pulsar. In their models, growth rates of
the electrostatic waves play an important role. Considering the plasma composed
from the thermal background plasma and hot and rare component with the
Dory-Guest-Harris distribution, we compute the growth rates $\gamma$ and
dispersion branches of the electrostatic waves in the $\omega - k_\perp$
domain. We show complexity of the electrostatic wave branches in the
upper-hybrid band. In order to compare the results, which we obtained
using the kinetic theory and Particle-in-cell (PIC) simulations, we define and
compute the integrated growth rate $\Gamma$, where the "characteristic width"
of dispersion branches was considered. We found a very good agreement between
the integrated growth rates and those from PIC simulations. For maximal and
minimal $\Gamma$ we showed locations of dispersion branches in the $\omega -
k_\perp$ domain. We found that $\Gamma$ has a maximum when the dispersion
branches not only cross the region with high growth rates $\gamma$, but when
the dispersion branches in this region are sufficiently long and wide. We also
mentioned the effects of changes in the background plasma and hot component
temperatures.
\end{abstract}


\section{Introduction}

Solar radio zebras belong to the most important fine structures used in
diagnostics of solar flare plasmas
\citep{2011fssr.book.....C,2011ApJ...736...64C,2013SoPh..284..579Z}. Similarly,
the zebras observed in the radio emission of Jupiter and Crab Nebula pulsar
\citep{2007ApJ...670..693H, 2016ApJ...833...47H,2018A&A...610A..69P} can also
be used for diagnostics purposes. Among many models of all these zebras, the
model based on the double plasma resonance (DPR) instability belongs to the
most probable \citep{1975SoPh...44..461Z,1982ApJ...259..844M,
1983SoPh...88..297Z,1986ApJ...307..808W,2001SoPh..202...71L,
2013SoPh..284..579Z,2015A&A...581A.115K,2016JGRA..121.5307Z,2018A&A...618A..60K,
2018ApJ...867...28K}.

In this model the DPR instability generates the upper-hybrid waves with the
frequency
\begin{equation}
\omega^2 = \omega_\mathrm{pe}^2 + \omega_\mathrm{ce}^2 +  3 k_\perp^2 v_\mathrm{tb}^2,
\label{eq0}
\end{equation}
when the resonance condition
\begin{equation}
\omega - \frac{k_\parallel u_\parallel}{\gamma_\mathrm{rel}} - \frac{s \omega_\mathrm{ce}}{\gamma_\mathrm{rel}} = 0,
\label{eq1}
\end{equation}
is fulfilled. Here $\omega$, $\omega_\mathrm{pe}$ , and
$\omega_\mathrm{ce}$ are the wave, electron-plasma, and
electron-cyclotron frequency, $\mathbf{k} = (k_\parallel, k_\perp)$ is wave
vector, $v_\mathrm{tb}$ the thermal electron velocity of the background plasma,
$\mathbf{u} = (u_\perp, u_\parallel)$, $u_\perp = p_\perp/m_\mathrm{e}$ , and
$u_\parallel = p_\parallel/m_\mathrm{e}$ are the hot electron velocities
perpendicular and parallel to the magnetic field; $m_\mathrm{e}$ is the
electron mass, $\gamma_\mathrm{rel}$ is the relativistic Lorentz factor, $s$ is the
gyro-harmonic number, and $c$ is the speed of light. For details, see
e.g.~\cite{2018A&A...611A..60B}.

In theoretical models of the double plasma resonance instability, a
two-component plasma  with the background plasma having the Maxwellian
distribution and hot and rare component with the Dory-Guest-Harris electron
distribution for $j=1$ \citep{1965PhRvL..14..131D}
\begin{equation}
f_\mathrm{hot}(u_\parallel, u_\perp) = \frac{u_\perp^2}{2 (2\pi)^{3/2} v_\mathrm{t}^5} \exp \left(-\frac{u_\perp^2 + u_\parallel^2}{2 v_\mathrm{t}^2}\right),
\label{eq2}
\end{equation}
is assumed. Here $v_\mathrm{t}$ we call the thermal velocity of hot electrons,
although the distribution function in this relation is not Maxwellian.

In interpretations of zebra observations, which use this model, it is usually
supposed that the electron-plasma frequency is related to the
electron-cyclotron frequency as follows
\begin{equation}
\omega_\mathrm{pe} \approx s \omega_\mathrm{ce}.
\end{equation}
However, \citet{2017A&A...555A...1B} showed that there can be the frequency
shift from this simple equation up to 16~\% for Dory-Guest-Harris velocity
distribution and even higher for other velocity distributions
\citep{2017SoPh..292..163Y}.

Besides the zebra model based on DPR instability, there is the model that
explains zebras by a nonlinear interaction of Bernstein modes
~\citep{2005A&A...438..341K,2009R&QE...52...88Z}. This model was supported by
zebra observations made by \citet{2005A&A...431.1037A}. In principle, Bernstein
modes can also be generated in the upper-hybrid band. Both the upper-hybrid
waves and Bernstein mode are the electrostatic (longitudinal) waves. Because in
the upper-hybrid band in some cases it is difficult to distinguish these waves,
therefore in the following we use the general term - the electrostatic waves.

As will be shown in the following, in the upper-hybrid band (i.e., at
frequencies close to the upper-hybrid frequency) there can be several
dispersion branches of the unstable electrostatic waves which energy can grow
simultaneously. Therefore, we define the "characteristic width" of
dispersion branches and compute the wave growth rates integrated over the
upper-hybrid band. Thus, for the first time we compare the results obtained
using the analytical kinetic theory with the results of PIC simulations. We
study the integrated growth rate in dependance on the ratio of the
electron-plasma and electron-cyclotron frequency and in relation to positions
of the wave dispersion branches in the $\omega - k_\perp$ domain. We use
3-dimensional PIC model. Note that PIC models naturally give the integrated
growth rates.

The paper is structured as follows. In Section 2 we start with a theory of the
electrostatic (longitudinal) waves perpendicular to the magnetic field. The
integrated growth rates of these waves in the upper-hybrid band computed from
the analytical relations are in Section~3. In Section 4 there are the growth
rates obtained numerically. Discussion of the results and conclusions are in
Sections 5.

\section{Electrostatic waves in the upper-hybrid band and their growth rates}

Let us consider the plasma composed from the background Maxwellian plasma with
the density $n_\mathrm{e}$ and the thermal velocity $v_\mathrm{tb}$ and hot
plasma having the density $n_\mathrm{h}$ and the "thermal" velocity
$v_\mathrm{t}$, where $n_\mathrm{h} \ll n_\mathrm{e}$. Then the dispersion
relation of the electrostatic (longitudinal) waves in such a plasma is given by
the permitivity tensor
\begin{equation}
\epsilon_\parallel = \epsilon_\parallel^{(0)} + \epsilon_\parallel^{(1)} = 0,
\label{eq4}
\end{equation}
where the term $\epsilon_\parallel^{(0)}$ corresponds to the background
Maxwellian plasma and the term $\epsilon_\parallel^{(1)}$ is a correction to
the hot and rare plasma component.

In our case with $n_\mathrm{h} \ll n_\mathrm{e}$ and in agreement with
\cite{1974itpp.book.....C,1997riap.book.....Z,fitzpatrick} we can write

\begin{equation}
\epsilon_\parallel^{(0)} = 1 - 2\omega_\mathrm{pe}^2 \frac{e^{-\lambda}}{\lambda} \sum_{l=1}^{\infty} \frac{l^2 I_l(\lambda)}{\omega^2 - l^2 \omega_\mathrm{ce}^2 } = 0,
\label{eq50}
\end{equation}
\begin{equation}
\omega_\mathrm{pe}^2 = \frac{n_\mathrm{e} e^2}{m_\mathrm{e}
\epsilon_\mathrm{0}}, \qquad \lambda = \frac{k_\perp^2
v_\mathrm{tb}^2}{\omega_\mathrm{ce}^2},
\end{equation}
where $\omega_\mathrm{pe}$ and $\omega_\mathrm{ce}$ is the plasma frequency
of the background plasma and the electron cyclotron frequency,
$\epsilon_\mathrm{0}$ is the permittivity of free space, $\lambda$ is
the dimensionless parameter, $I_l(\lambda)$ is the modified Bessel function of
$l$th order, $m_\mathrm{e}$ is the electron mass, $e$ is the electron charge,
$\omega$ is the wave frequency, $\mathbf{k} = (k_\parallel, k_\perp)$ is the
wave vector parallel and perpendicular to the direction of the magnetic field,
respectively.

Solutions of the real part of Equation~\ref{eq50} are the dispersion relations
for the upper-hybrid waves as well as for the Bernstein waves. Because in the
present paper we are interested about these waves in the upper-hybrid band,
where sometimes is a problem to distinguish between these waves, therefore in
the following we use for them the common term: the electrostatic waves.

For the growth rate of these waves, we can write \citep{1975SoPh...43..431Z}
\begin{equation}
\gamma(\omega, k_\perp) = - \frac{ \mathrm{Im} \, \epsilon_\parallel^{(1)}}
{\left[\frac{\partial \mathrm{Re} \, \epsilon_\parallel^{(0)}}
{\partial \omega}\right]_{\epsilon_\parallel^{(0)} = 0}}.
\label{eq6}
\end{equation}

In accordance with \citet[Appendix A]{2005A&A...438..341K} the nominator of
Equation~\ref{eq6} can be written as
\begin{equation}
\mathrm{Im}(\epsilon_\parallel^{(1)}) = - 2\pi^2 m_\mathrm{e}^4  \frac{\omega_\mathrm{pe}^2}{k^2}
    \sum_{l=s+1}^{\infty} ab^2 \int_0^{\pi}  J_l \left( \frac{\gamma_\mathrm{rel} k_\perp v_\perp}{\omega_\mathrm{ce}} \right) \frac{\gamma_\mathrm{rel}^5 \sin \phi}{\frac{\partial \psi}{\partial\rho}}  \frac{l\omega_\mathrm{ce}}{\gamma_\mathrm{rel} v_\perp}\frac{\partial f}{\partial p_\perp} \, \mathrm{d}\phi,
\label{eq7}
\end{equation}
\begin{equation}
\frac{\partial \psi}{\partial \rho} = \frac{\gamma_\mathrm{rel}^2 l \omega_\mathrm{ce}}{c^2} \left( v_\parallel^2 + v_\perp^2 \right),
\end{equation}

\begin{equation}
    v_\parallel = - a \cos(\phi), \quad v_\perp = b \sin(\phi),
    \label{eq9}
\end{equation}

\begin{equation}
    a^2 = \frac{l^2 \omega_\mathrm{ce}^2 c^2 (l^2 \omega_\mathrm{ce}^2 - \omega^2)}
    {l^4 \omega_\mathrm{ce}^4},
    \label{eq11}
\end{equation}

\begin{equation}
b^2 = \frac{c^2 (l^2 \omega_\mathrm{ce}^2 - \omega^2)}{l^2 \omega_\mathrm{ce}^2},
\label{eq12}
\end{equation}
where $c$ is the speed of light, $v_\parallel$ and $v_\perp$ are the velocities
on resonance ellipse in Equation~\ref{eq1}. $f$ is the electron velocity
distribution function in the form
\begin{equation}
f(v_\perp,v_\parallel) = f_0(v_\perp,v_\parallel) + \frac{n_\mathrm{h}}{n_\mathrm{e}} f_\mathrm{hot}(v_\perp,v_\parallel).
\label{eq9}
\end{equation}
The $f_0$ means the background Maxwellian distribution for temperature
$v_\mathrm{tb}$, $f_\mathrm{hot}$ is in our case the Dory-Guest-Harris
distribution given by Equation~\ref{eq2} described by the "thermal" velocity
$v_\mathrm{t}$ and $J_l(\lambda)$ is the Bessel function.

Furthermore, the denominator of Equation~\ref{eq6} can be expressed as

\begin{equation}
\frac{\partial \epsilon_\parallel}{\partial \omega} = 4 \omega \omega_\mathrm{pe}^2 \frac{e^{-\lambda}}{\lambda} \sum_{l=1}^\infty \frac{l^2 I_l(\lambda)}{(\omega^2 - l^2 \omega_\mathrm{ce}^2)^2}.
\label{eq10}
\end{equation}

\section{Analytical growth rates}

In this Section, we compute the growth rates using analytically derived
equations expressed in the previous Section. As an example we make computations
for the ratio of the electron-plasma and electron-cyclotron frequency
($\omega_\mathrm{pe}/\omega_\mathrm{ce}$) in the 4.0-5.3 range. Namely, we want
to determine the growth rates also for a non-integer ratio of
$\omega_\mathrm{pe}/\omega_\mathrm{ce}$. The relatively low values of this
ratio are chosen due to a comparison with the numerical simulations, where
computations with the low values of $\omega_\mathrm{pe}/\omega_\mathrm{ce}$ are
more reliable.

If we do not mention explicitly, in our analytical computations and also in the
following numerical simulations, we use the parameters shown in
Table~\ref{tab1}.
Considering the propagation of waves in the strictly perpendicular direction to
the magnetic field ($k_\parallel=0$) and using Equations~\ref{eq50}
we compute their dispersion branches and growth rates in the
$\omega - k_\perp$ domain.

Because we want the frequency precision of dispersion branches at least
$10^{-8} \omega_\mathrm{pe}$, we searched for a sufficient number $l$ in
relation~\ref{eq50}. We increased $l$ until the precision of a dispersion
branch position was higher than $10^{-8} \omega_\mathrm{pe}$. Thus, we got
$l_\mathrm{max} = 40$ and in relation~\ref{eq50} we use the summation
$\sum_{l=1}^{l_\mathrm{max} = 40}$. The same number $l$ is also used in the
summation in Equation~\ref{eq7}.

\begin{table}
\centering
\begin{tabular}{c|c}
\hline\hline
Parameter & Value \\
\hline
$v_\mathrm{tb}$ & 0.018~$c$ (2~MK) \\
$v_\mathrm{t}$ & 0.2~$c$\\
$n_\mathrm{e}/n_\mathrm{h}$ & 32 \\
$f_\mathrm{hot}$ & DGH \\
$\omega_\mathrm{pe}/\omega_\mathrm{ce}$ & $4.0-5.3$ \\
$k_\perp c / \omega_\mathrm{pe}$ & $0-15$ range \\
$\omega / \omega_\mathrm{pe}$ & $0-2$ range\\
\hline
\end{tabular}
\caption{Parameters used for computing of the growth rates.} \label{tab1}
\end{table}

Roots of Equation~\ref{eq50} (dispersion branches) are searched numerically
using the Levenberg--Marquardt damped root method
\citep{Levenberg,Marquardt,More:126569,Press:2007:NRE:1403886} from SciPy
package in the Python. This methods was selected for its good convergence
properties and ability to find the most narrower branches. The minimization is
made in variable $\omega$. First, the $\omega - k_\perp$ domain is divided into
the regular orthogonal grid, which gives starting values of $\omega$ and
$k_\perp$ for the algorithm. Typically, we use $10^3-10^4$ grid points in
$\omega$ direction and $20-40$ grid points in $k_\perp$. In each step the
method computes gradients in the Jacobian matrix and by the gradient descent
method it proceeds until the required precision of $\omega$ is
obtained. In our case we set this precision as $10^{-8} \omega_\mathrm{pe}$.
From the last step in this procedure we get the covariance matrix, which
represents the gradient in a neighborhood of the found solution. We take the
inverted absolute value of this gradient as the "characteristic width"
of the dispersion branches, see the following. From physical point of view the
"characteristic width" of the dispersion branch is given by thermal
fluctuations of the electron plasma density. This "characteristic
width" does not represent the actual relative frequency range occupied by the
waves, but is rather proportional to that frequency range; the proportionality
coefficient depends on many factors (fluctuations etc.), but its absolute value
is unimportant for this study.

\begin{figure}[tbh!]
    \centering
    \includegraphics[width = 8.cm]{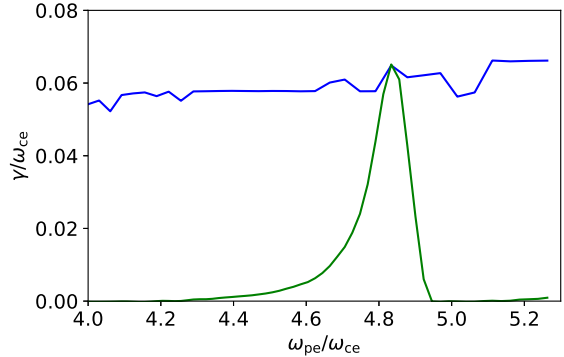}
    \caption{The maximal growth rate found at the upper-hybrid branch (Equation~\ref{eq0}) (green line),
    and at all electrostatic branches (Equation~\ref{eq50}) (blue line).
    The frequency and wave vector intervals are the same as for the integrated growth rate in Figure~\ref{fig1}.}
    \label{fig1c}
\end{figure}

First, we compute the maxima of the growth rate $\gamma$ at the
upper-hybrid branch (Equation~\ref{eq0}) in dependance on
$\omega_\mathrm{pe}/\omega_\mathrm{ce}$. The result is shown in
Figure~\ref{fig1c} by the green line. Then using Equation~\ref{eq50} we compute
the maximal growth rate at all electrostatic branches in the $\omega - k_\perp$
domain, see the blue line in Figure~\ref{fig1c}. Note that these maximal growth
rates are always taken in one specific point of the $\omega - k_\perp$ domain.
While the growth rate at the upper-hybrid branch has distinct maximum
($\gamma/\omega_\mathrm{ce} \approx$ 0.06) at about
$\omega_\mathrm{pe}/\omega_\mathrm{ce}$ = 4.8, the growth rates for all
branches are similar. It means that at almost all branches in the $\omega -
k_\perp$ domain there are positive growth rates.

Because our main goal is to to compare the results obtained from
analytical relations with those from the following PIC simulations, where the
growth rate is given by compositions of growth rates in the whole $\omega -
k_\perp$ domain, we define a new quantity -- the integrated growth rate
$\Gamma$ as
\begin{equation}
\Gamma = \frac{1}{\Gamma_0} \int \gamma(\omega, k_\perp) \sigma(\omega, k_\perp) \delta(\epsilon_\parallel^{(0)}(\omega, k_\perp)) \mathrm{d}\omega \mathrm{d}k_\perp,
\label{eq11}
\end{equation}
where $\delta$ is the Dirac delta function and $\gamma(\omega, k_\perp)$ is the
growth rate at the specific $\omega$ and $k_\perp$. This integral counts the
growth rates over the dispersion branches of the  electrostatic waves in the
the whole $\omega - k_\perp$ domain with their "characteristic width"
$\sigma$. The function $\Gamma_0$ is the normalization factor
\begin{equation}
\Gamma_0 = \int \sigma(\omega, k_\perp) \delta(\epsilon_\parallel^{(0)}(\omega, k_\perp)) \mathrm{d}\omega \mathrm{d}k_\perp.
\label{eq12}
\end{equation}
In our case the integral is computed over the $\omega - k_{\perp}$ area with
$\omega/\omega_\mathrm{pe} \in (0,2)$ and $k_\perp c / \omega_\mathrm{pe} \in
(0,20)$. At higher frequencies the branches are very narrow and thus do not
contribute effectively to the integrated growth rate and at higher values of
the wave vectors $\gamma = 0$.

\begin{figure}[tbh!]
   \centering
  \includegraphics[width = 8.cm]{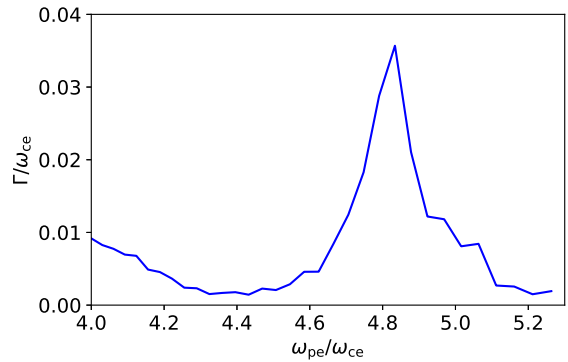}
  \caption{Integrated growth rate $\Gamma$ computed from Equation~\ref{eq11} as a function
  of the ratio $\omega_\mathrm{pe}/\omega_\mathrm{ce}$ normalized to $\omega_\mathrm{ce}$ for the parameters given in Table~\ref{tab1}.
  Its maximum is for $\omega_\mathrm{pe}/\omega_\mathrm{ce} = 4.8$ and minimum for $\omega_\mathrm{pe}/\omega_\mathrm{ce} = 4.4$.}
  \label{fig1}
\end{figure}

Using the plasma parameters from Table~\ref{tab1} we computed the integrated
growth rates $\Gamma$ (Equation~\ref{eq11}) for the ratios
$\omega_\mathrm{pe}/\omega_\mathrm{ce}$ = 4.0 - 5.3, see Figure~\ref{fig1}. As
can be seen in this figure, the maximum of $\Gamma$ is for
$\omega_\mathrm{pe}/\omega_\mathrm{ce}$ = 4.8 and the minimum for
$\omega_\mathrm{pe}/\omega_\mathrm{ce}$ = 4.4.

\begin{figure*}
        \centering
        \gridline{\fig{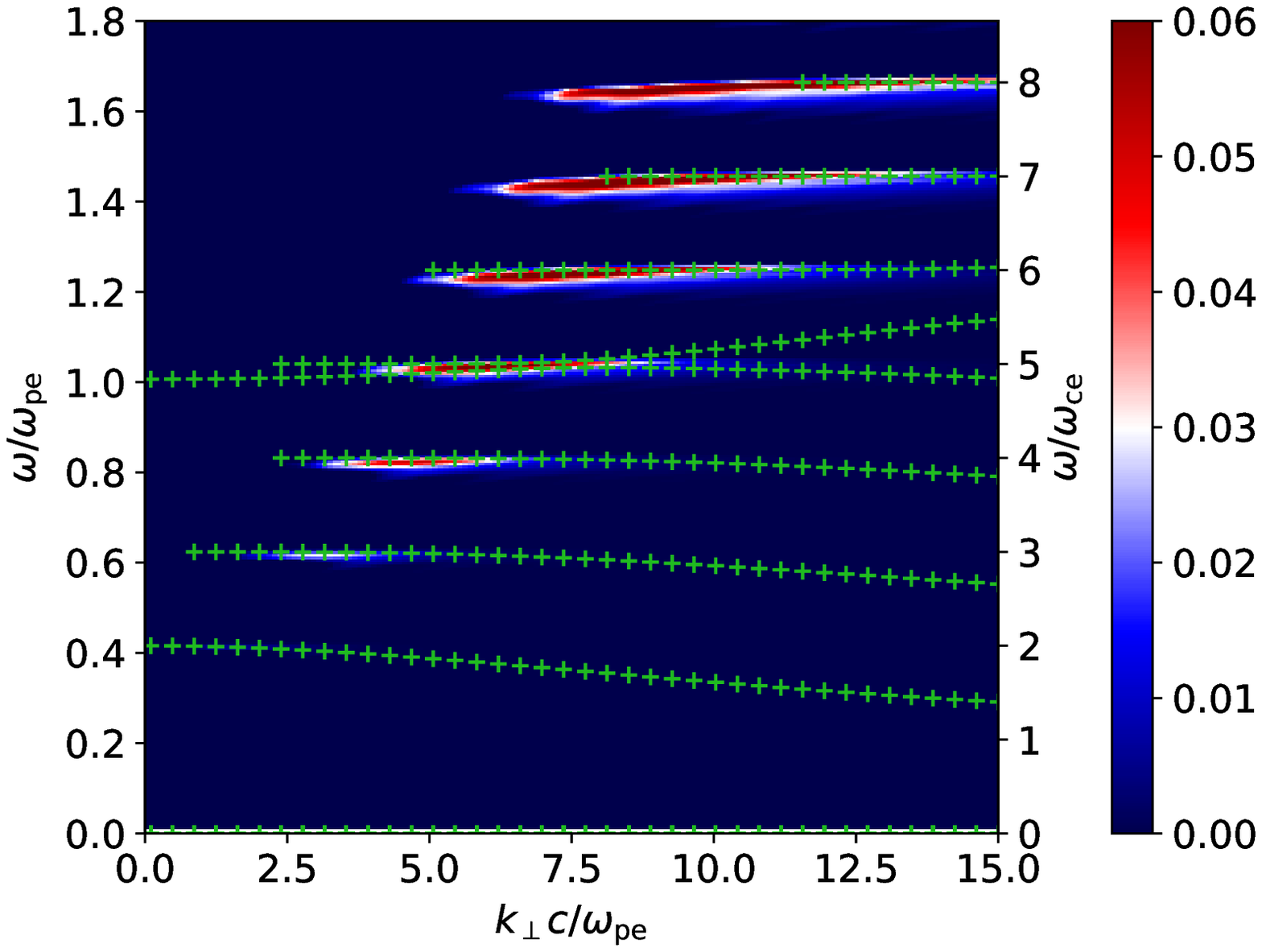}{0.49\textwidth}{(a)}
                  \fig{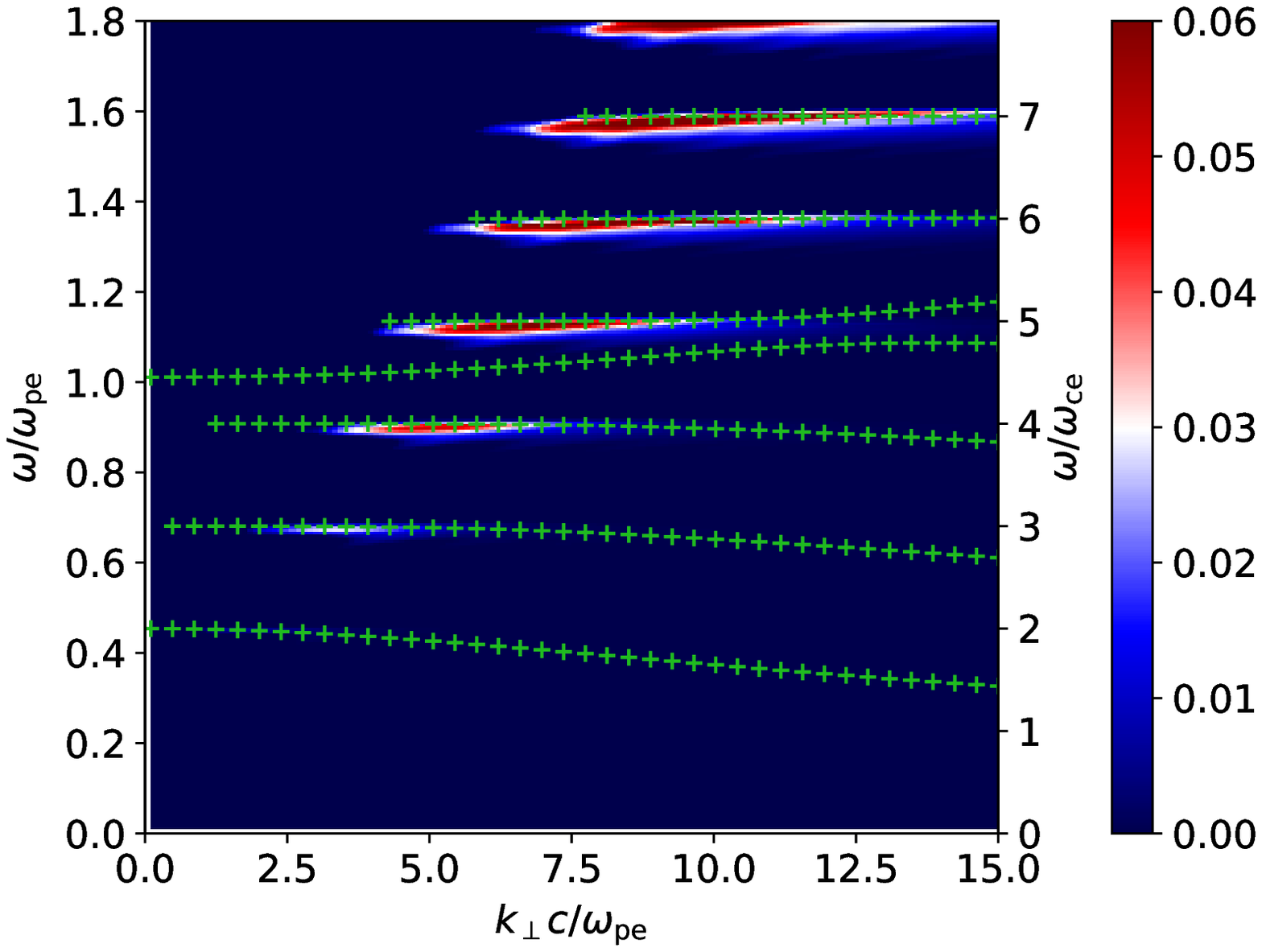}{0.49\textwidth}{(b)}
          }
        \gridline{
                  \fig{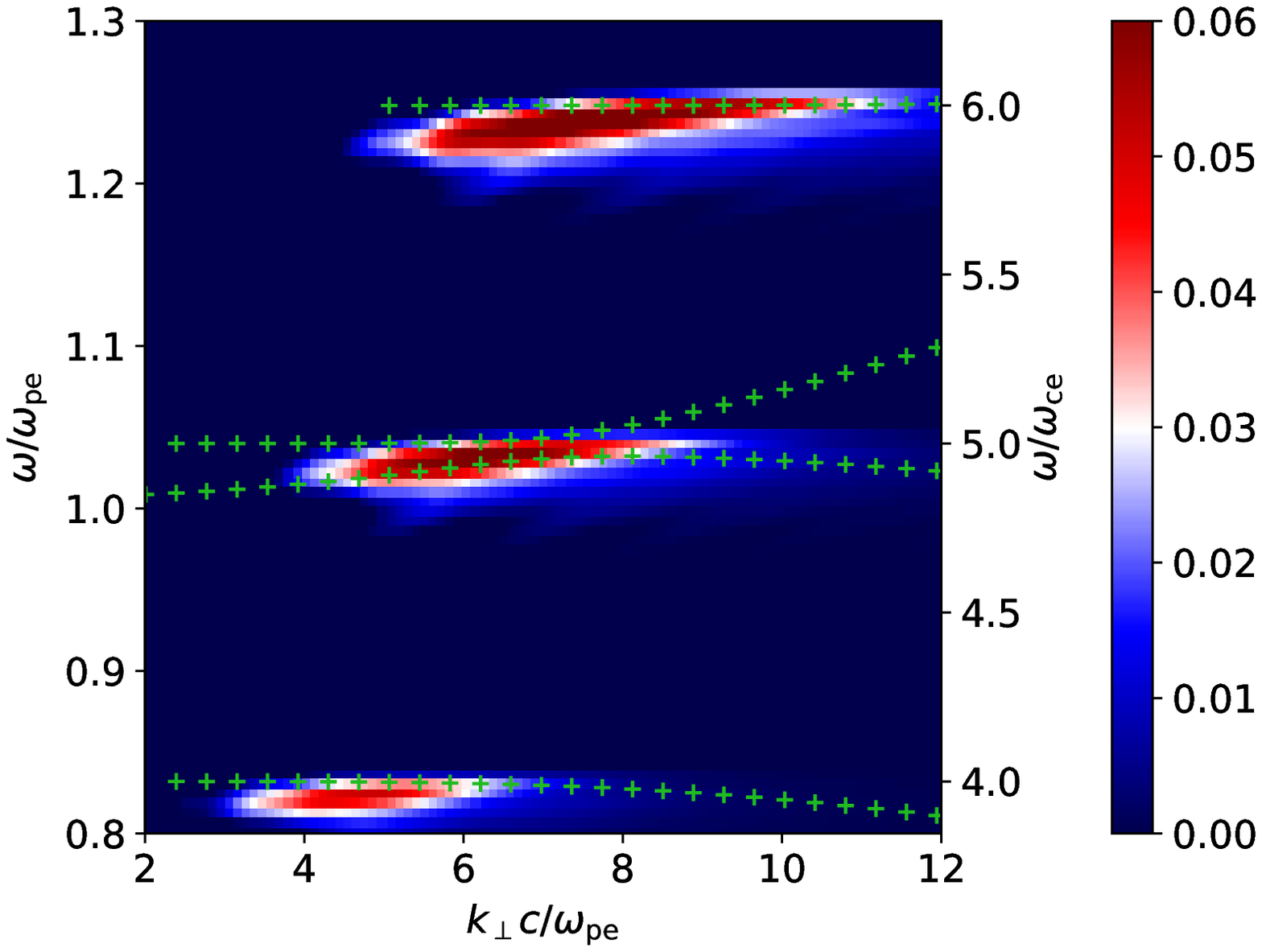}{0.49\textwidth}{(c)}
                  \fig{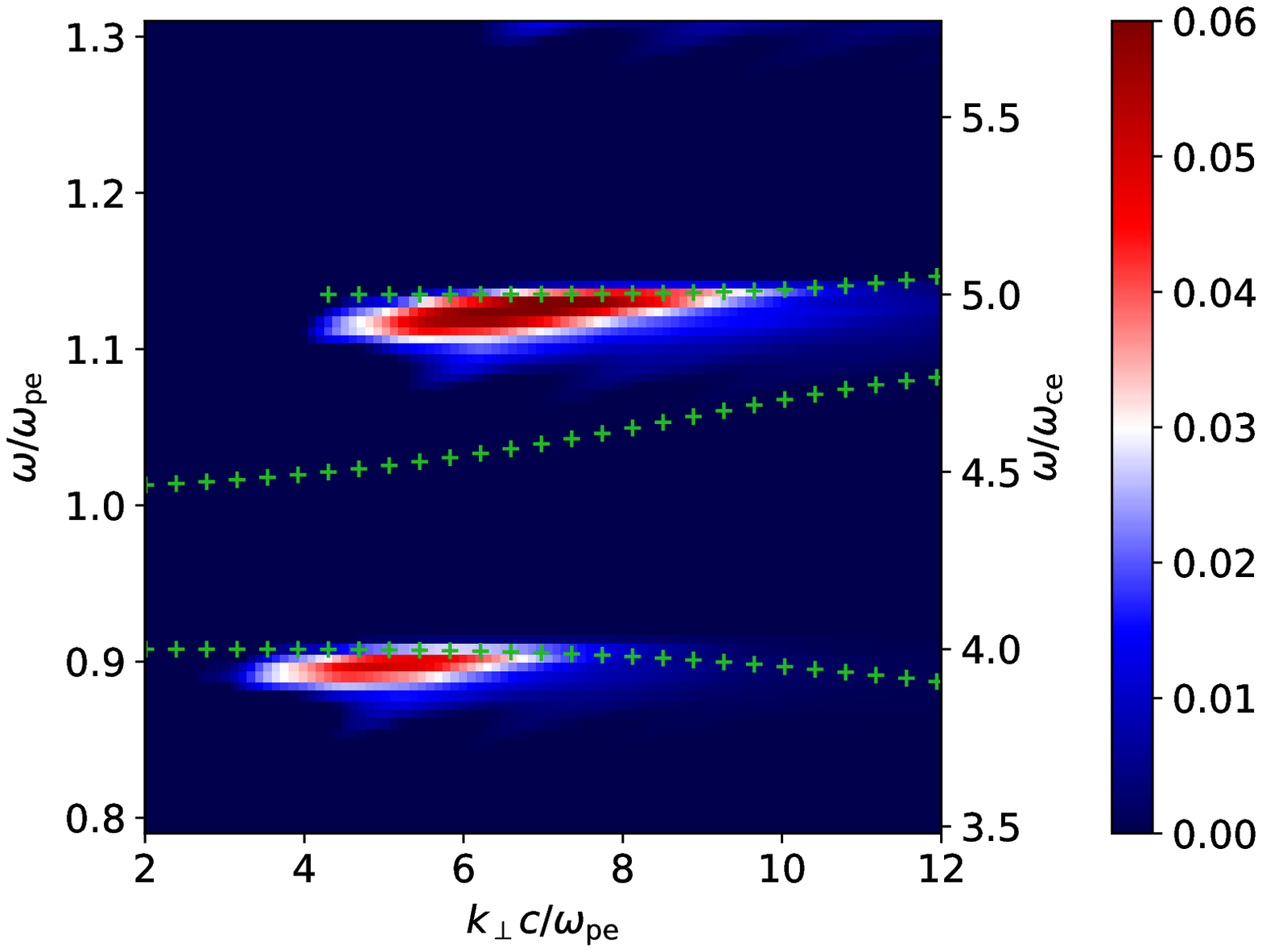}{0.49\textwidth}{(d)}
          }
    \caption{Analytical growth rates $\gamma/\omega_\mathrm{ce}$ and dispersion branches in the $\omega - k_\perp$ domain (the blue-red
    scale)
    for the maximal value of $\Gamma$, i.e., for $\omega_\mathrm{pe}/\omega_\mathrm{ce}=4.8$ (a, c) and for
    the minimal value of $\Gamma$, i.e., for $\omega_\mathrm{pe}/\omega_\mathrm{ce}=4.4$ (b, d) taken from Figure~\ref{fig1}. The green crosses show dispersion branches
    of the electrostatic waves. The growth rates and dispersion branches in a broad range of $(\omega,k_\perp)$ are in (a, b) and
    the same in a detailed view in the upper-hybrid band are in (c, d).}
\label{fig2}
\end{figure*}

Now, a question arises how the integrated growth rate $\Gamma$ is related to
the positions of dispersion branches in the $\omega - k_\perp$ domain. To
answer this question, we compute the dispersion branches in the $\omega -
k_\perp$ domain for the maximal and minimal values of the integrated growth
rate $\Gamma$. We also plot growth rates in the whole domain $\omega -
k_\perp$, although growth rates outside dispersion branches are non-physical.
It is because the growth rates are defined only for roots of
Equation~\ref{eq50}. But, we use this type of presentation to show a relation
between positions of dispersion branches and regions favorable for the
wave amplification.

The results are shown in Figure~\ref{fig2}, where the growth rates and
dispersion branches are in a broad area of $\omega - k_\perp$ (Plots a, b). The
detailed views are from the upper-hybrid band (Plots c, d). We note that the
parts of dispersion branches with the "characteristic width" $\sigma$
less than 10$^{-6}$ are very narrow and thus they are not represented in the
figure, see also the following and Table~\ref{tab2}.

As seen here, in the $\omega - k_\perp$ domain there are the regions with the
high growth rates. The maximal growth rate on dispersion curves in the field of
view of Figure~\ref{fig2}c is $\gamma/\omega_\mathrm{ce} \approx 0.06$.
Comparing the cases with the maximal and minimal integrated growth rates
$\Gamma$, we can see differences in distributions of the dispersion branches
and growth rate regions. While in the case with the maximal $\Gamma$ the
dispersion branches in most cases cross the region with the high growth rates
$\gamma$, in the case with the minimal $\Gamma$ the thickest branches are out
these regions. Note that for the integrated growth rate it is important not
only this crossing over these regions but also the lengths and width (area) of
the dispersion branches over these regions. Namely, the wave energy is given by
the wave energy density in the area unit times the area. We suppose that the
width of the dispersion branch is proportional to $\sigma$ calculated during
computations of the dispersion branches. Thus, there are the dispersion
branches, which go through the high growth rate regions even for the minimal
growth rate $\Gamma$ (see Figure~\ref{fig2}b, e.g. the region
$\omega/\omega_\mathrm{pe}=1.4$, $k=6-15$), but their "characteristic
width" is very small and therefore do not significantly influence the
integrated growth rate $\Gamma$.

\begin{table}
\centering
\begin{tabular}{ccc|ccD}
\hline\hline
\multicolumn{3}{c|}{Maximum} & \multicolumn{3}{c}{Minimum} \\
\multicolumn{3}{c|}{$\omega_\mathrm{pe}/\omega_\mathrm{ce} = 4.8$} & \multicolumn{3}{c}{$\omega_\mathrm{pe}/\omega_\mathrm{ce} = 4.4$} \\
$\omega_\mathrm{branch}/\omega_\mathrm{pe}$ & $\omega_\mathrm{branch}/\omega_\mathrm{ce}$ & $\sigma$ & $\omega_\mathrm{branch}/\omega_\mathrm{pe}$ & $\omega_\mathrm{branch}/\omega_\mathrm{ce}$ & $\sigma$  \\
\hline
 0.334 & 1.603 & $7.4\times 10^{-3}$ & 0.373 & 1.788 & $9.1\times 10^{-3}$  \\
 0.593 & 2.846 & $1.0 \times 10^{-2}$ & 0.652 & 3.128 & $1.2\times 10^{-2}$  \\
 0.821 & 3.943 & $1.2 \times 10^{-2}$ & 0.897 & 4.306 & $2.0 \times 10^{-2}$  \\
 1.030 & 4.945 & $9.9 \times 10^{-2}$ & 1.069 & 5.133 & $4.0 \times 10^{-1}$  \\
 1.076 & 5.164 & $3.4 \times 10^{-1}$ & 1.140 & 5.471 & $2.2 \times 10^{-2}$  \\
 1.250 & 6.001 & $6.6 \times 10^{-4}$ & 1.364 & 6.546 & $1.46 \times 10^{-4}$ \\
 1.456 & 6.989 & $1.5 \times 10^{-5}$ & 1.591 & 7.636 & $3.95 \times 10^{-6}$ \\

\hline
\end{tabular}
\caption{"Characteristic width" of the dispersion branches $\sigma$
for the maximum $\Gamma$ with $\omega_\mathrm{pe}/\omega_\mathrm{ce} = 4.8$
(see Figure~\ref{fig1}) and for the minimal $\Gamma$ with
$\omega_\mathrm{pe}/\omega_\mathrm{ce} = 4.4$. $\omega_\mathrm{branch}$ means
the frequency  of the dispersion branch for $k_\perp c /
\omega_\mathrm{pe}=10$.} \label{tab2}
\end{table}

In the Table~\ref{tab2} we present the "characteristic width" of
different dispersion branches $\sigma$. A difference in the
"characteristic width" can be several orders. The maximal $\Gamma$
appears, when the dispersion branch with the highest "characteristic
width" crosses the high growth rate area around the plasma frequency, see
Figure~\ref{fig2}c. It is interesting that in this case, two dispersion
branches of the electrostatic waves are very close to each other.

\begin{figure*}
    \centering
    \gridline{
            \fig{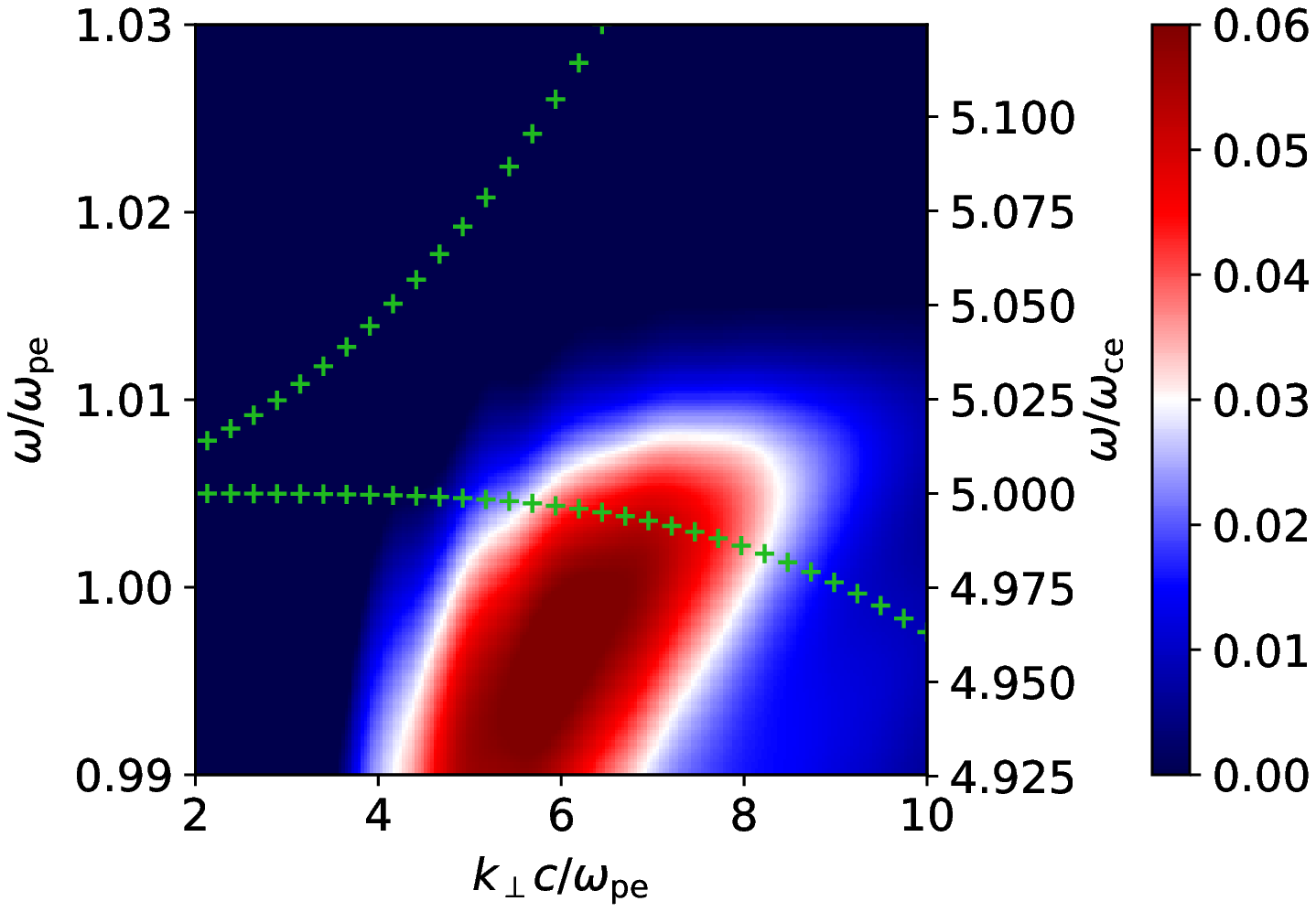}{0.45\textwidth}{(a)}
            \fig{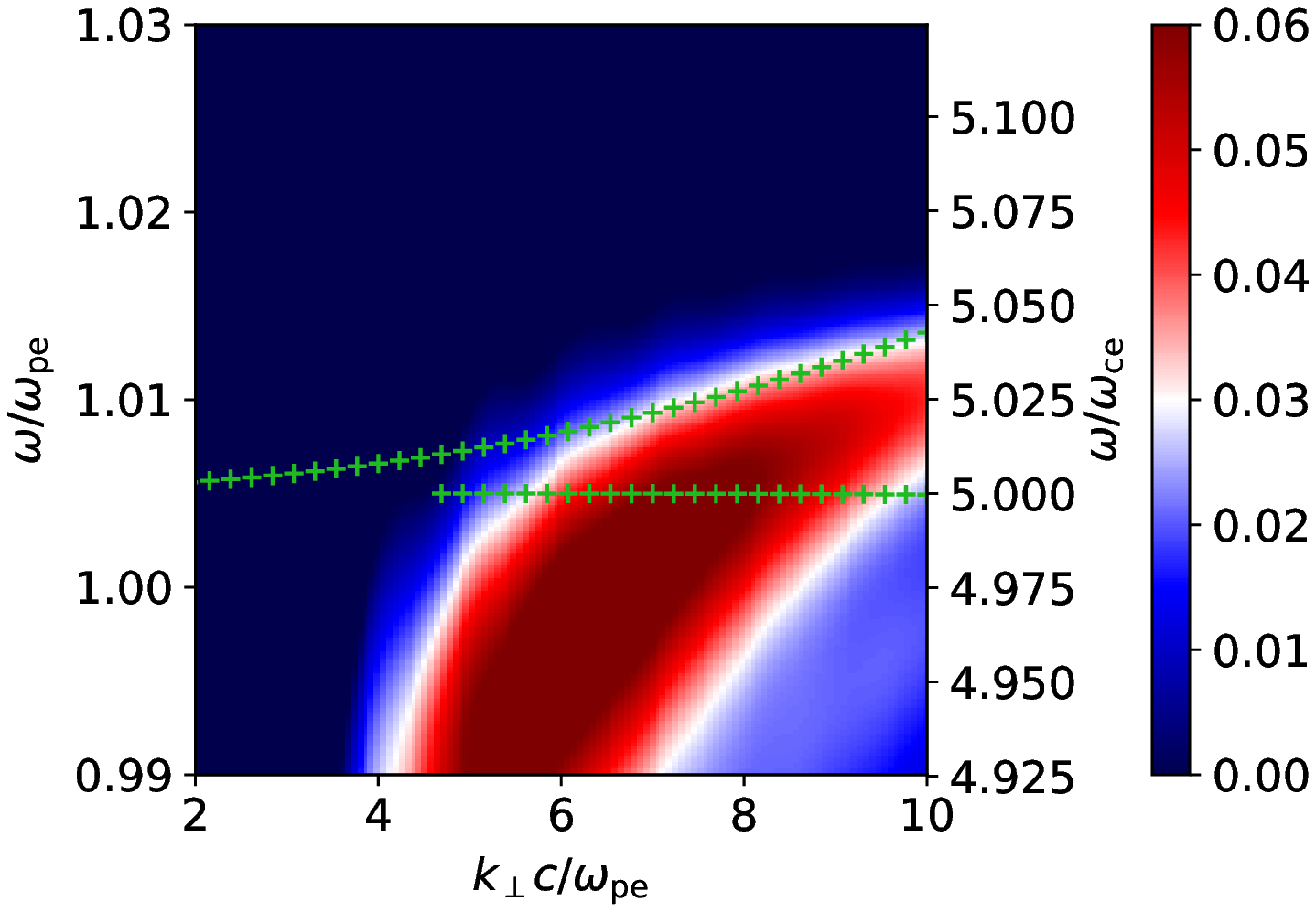}{0.45\textwidth}{(b)}
    }
    \gridline{
            \fig{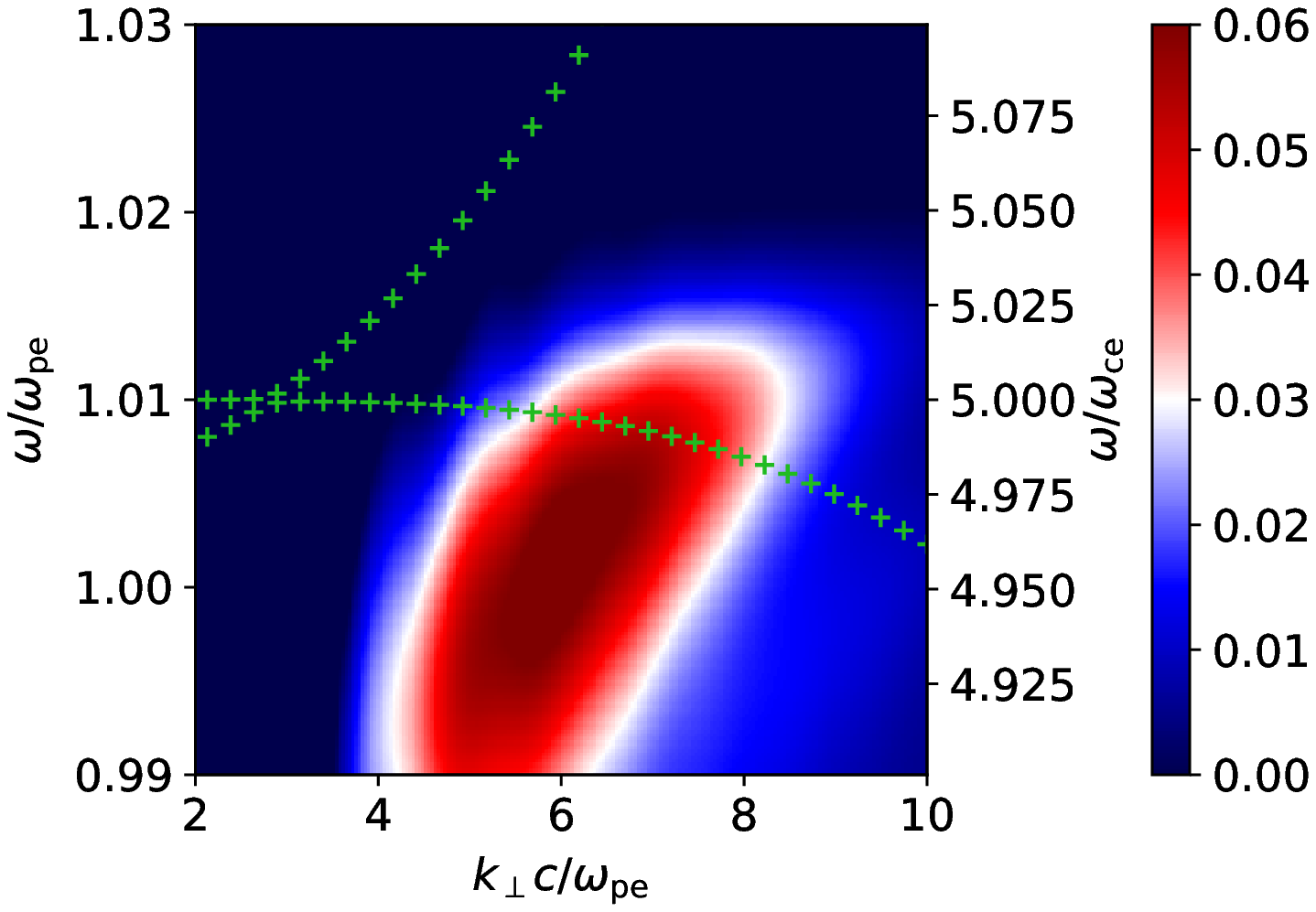}{0.45\textwidth}{(c)}
            \fig{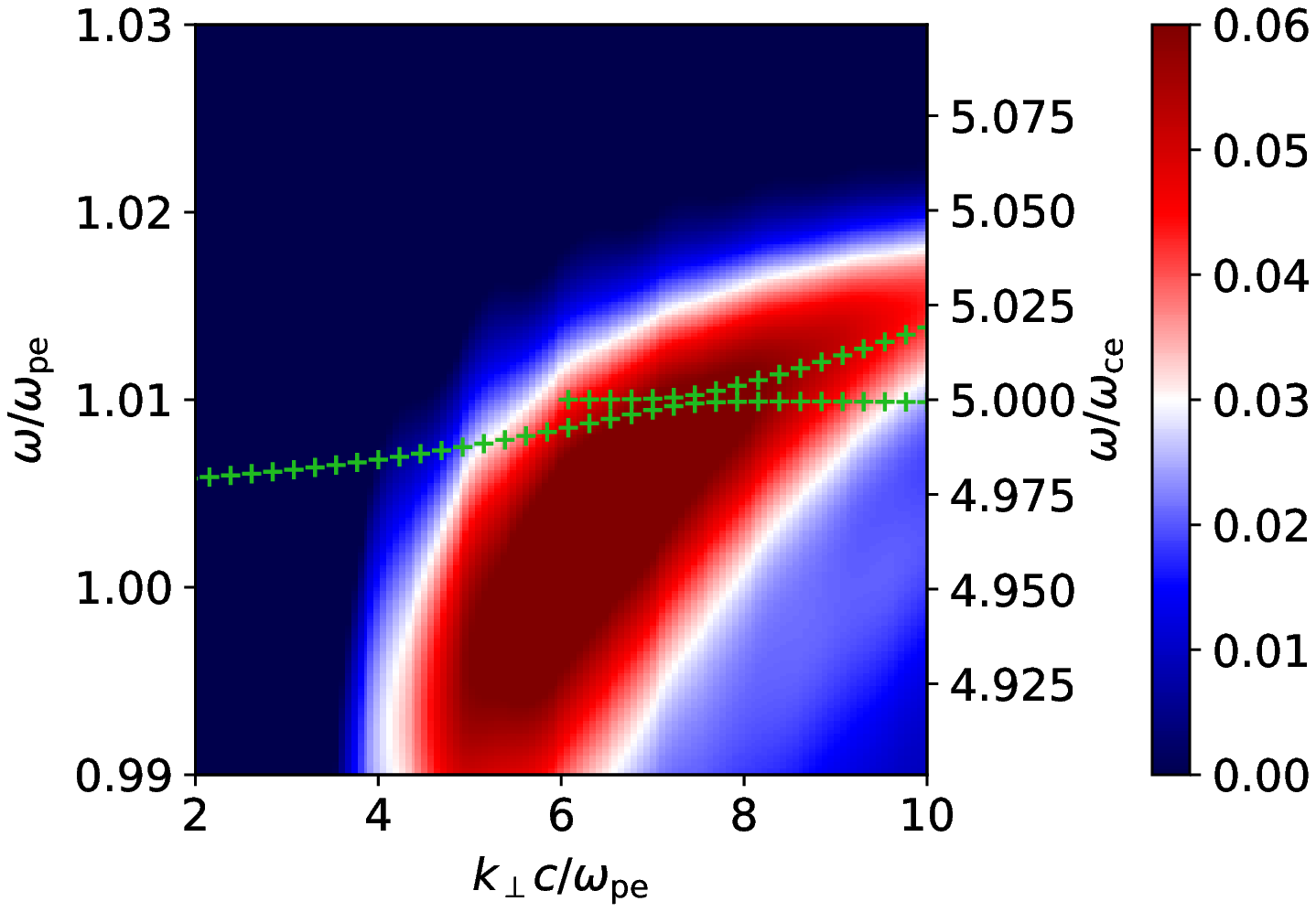}{0.45\textwidth}{(d)}
    }
    \gridline{
            \fig{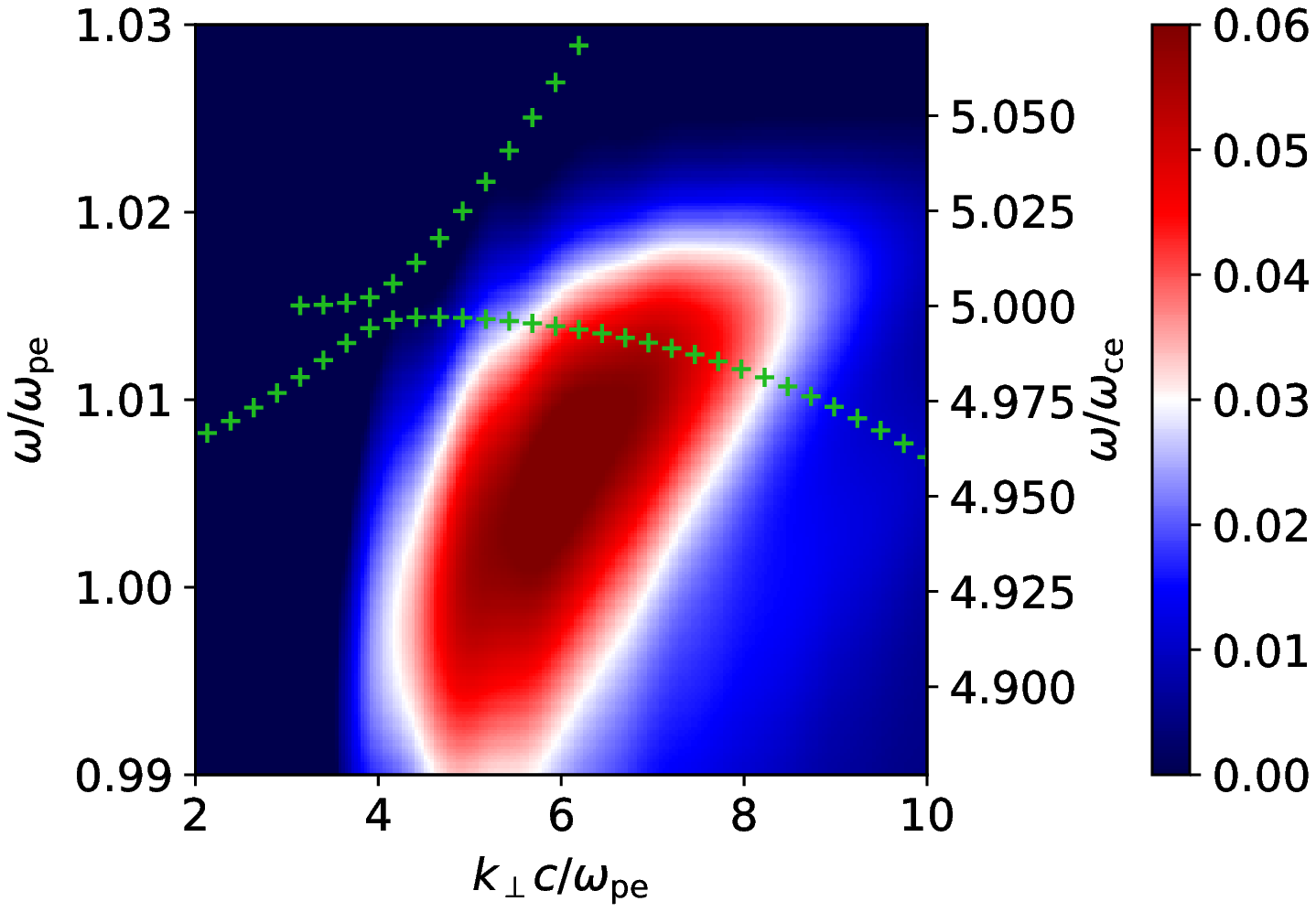}{0.45\textwidth}{(e)}
            \fig{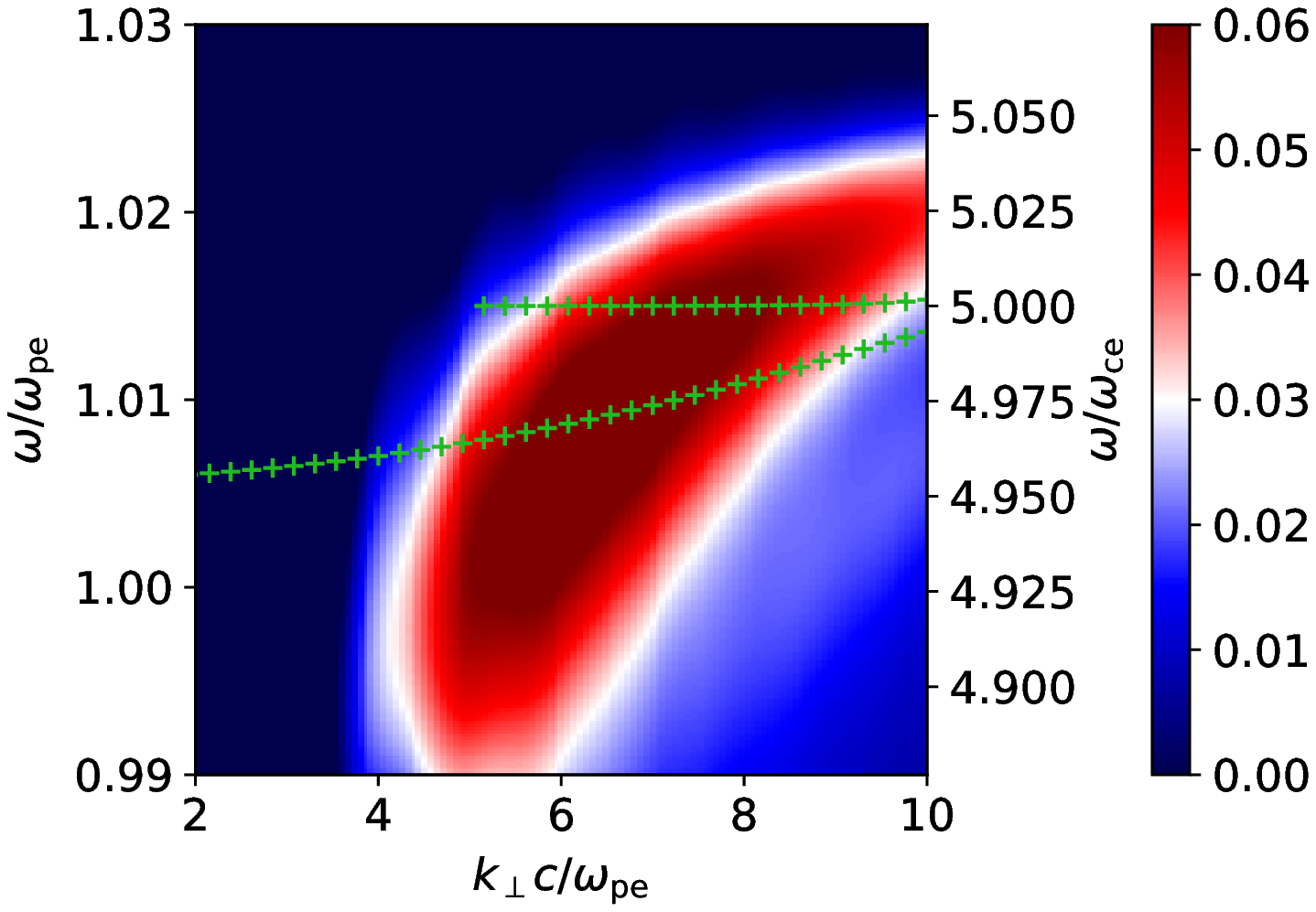}{0.45\textwidth}{(f)}
    }
    \caption{Analytical growth rates normalized to $\omega_\mathrm{ce}$ and dispersion branches in dependence
    on ratio $\omega_\mathrm{pe}/\omega_\mathrm{ce}$ and 4.975 (a, b), 4.950 (c, d), 4.925 (e, f)  for two background
    temperatures: $v_\mathrm{tb} = 0.018$~$c$ (a, c, e) and $v_\mathrm{tb} = 0.007$~$c$ (b, d, f).} \label{fig3}
\end{figure*}

To show how the dispersion branches change, in Figure~\ref{fig3} we present the
dispersion branches for three values of the ratio
$\omega_\mathrm{pe}/\omega_\mathrm{ce}$ = 4.925, 4.950 and 4.975. Here in the
left part of the figure we can see that the dispersion branch, which is firstly
under the plasma frequency, is going up to higher frequencies with decreasing
$\omega_\mathrm{pe}/\omega_\mathrm{ce}$. During this shift it extrudes the
upper dispersion branch up (Figure~\ref{fig3}c). In the region where the
branches meet a knee on the bottom dispersion branch is formed; i.e., for lower
$k_\perp$ than this knee there is a part of the dispersion branch with the
normal dispersion and for higher $k_\perp$ is the part with the anomalous
dispersion (Figure~\ref{fig3}c). Thus, in some cases, the electrostatic waves
are generated at the part with the normal dispersion and in others at that with
the anomalous dispersion.

For the value of $v_\mathrm{tb} = 0.018$~$c$ considered in
Figure~\ref{fig3}a,c,e, this interplay of dispersion branches happens slightly
out of the region with high growth rates; therefore the integrated growth rate
$\Gamma$ has not the maximal value. However, we found that for lower background
thermal velocity ($v_\mathrm{tb} = 0.007$~$c$), the dispersion branches not
only meet, but they also are in the region with high growth rates
(Figure~\ref{fig3}d, f) and thus for these plasma parameters the high
integrated growth rate $\Gamma$ can be expected.

\begin{figure*}
    \centering
    \gridline{
            \fig{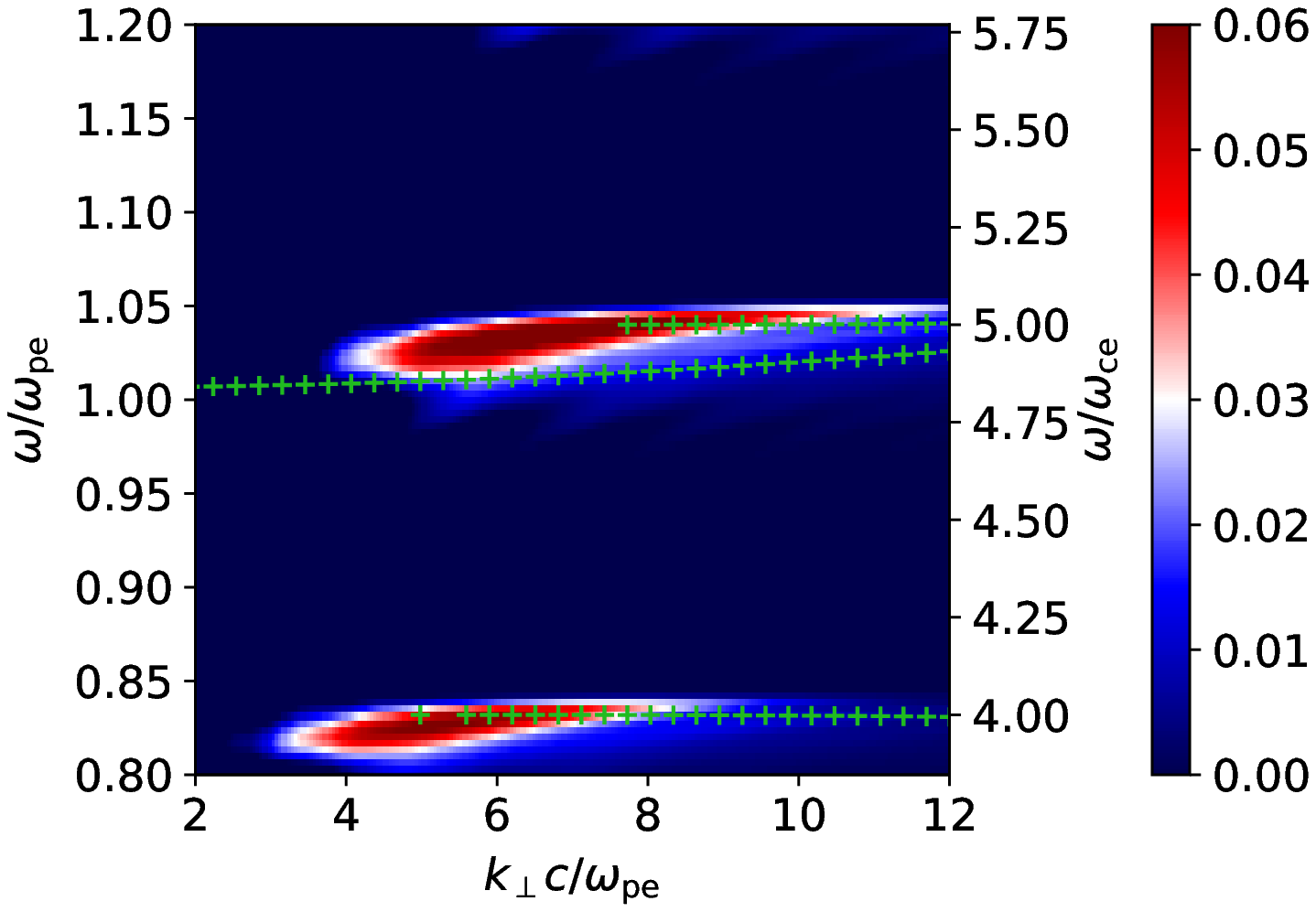}{0.32\textwidth}{(a)}
            \fig{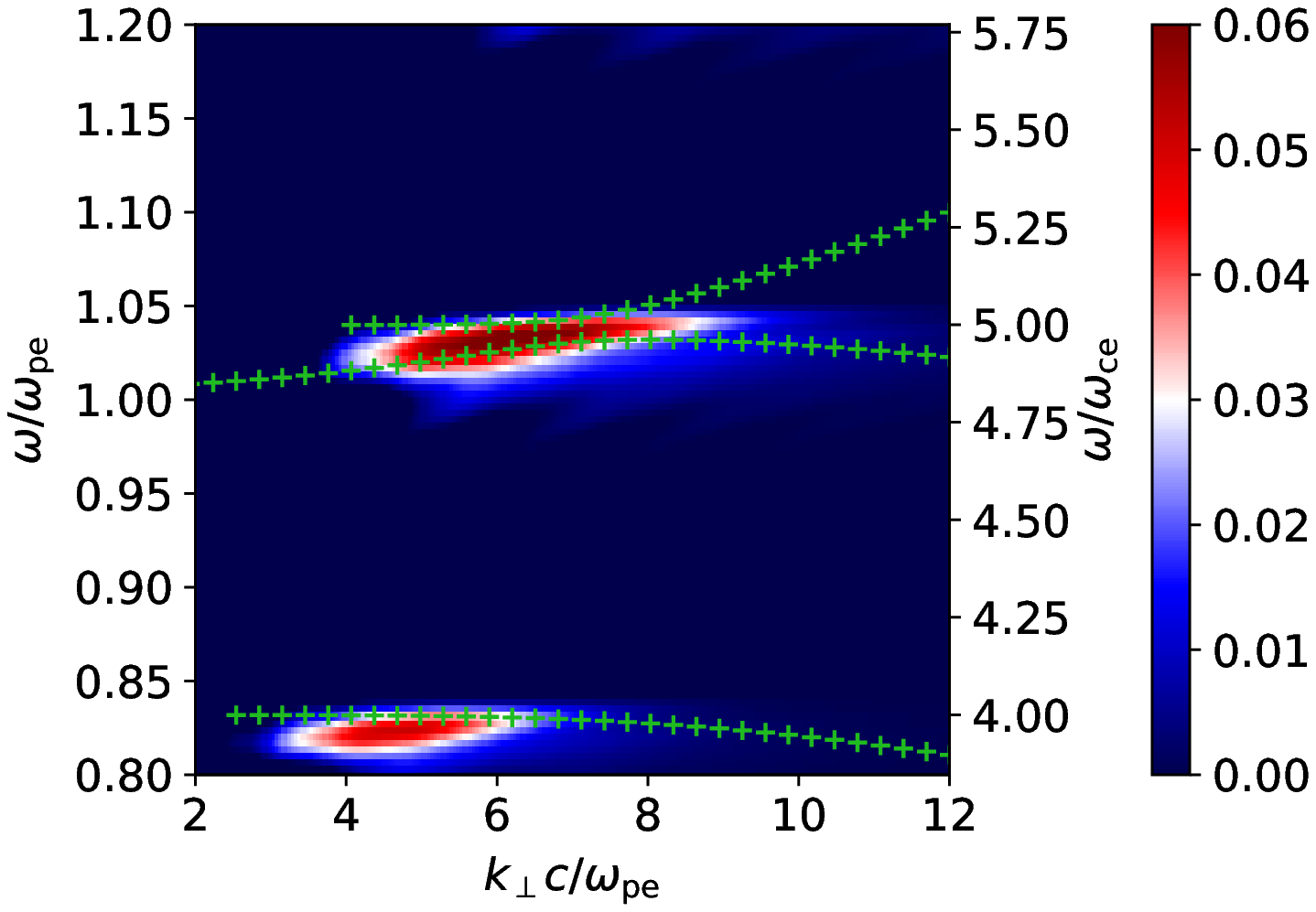}{0.32\textwidth}{(b)}
            \fig{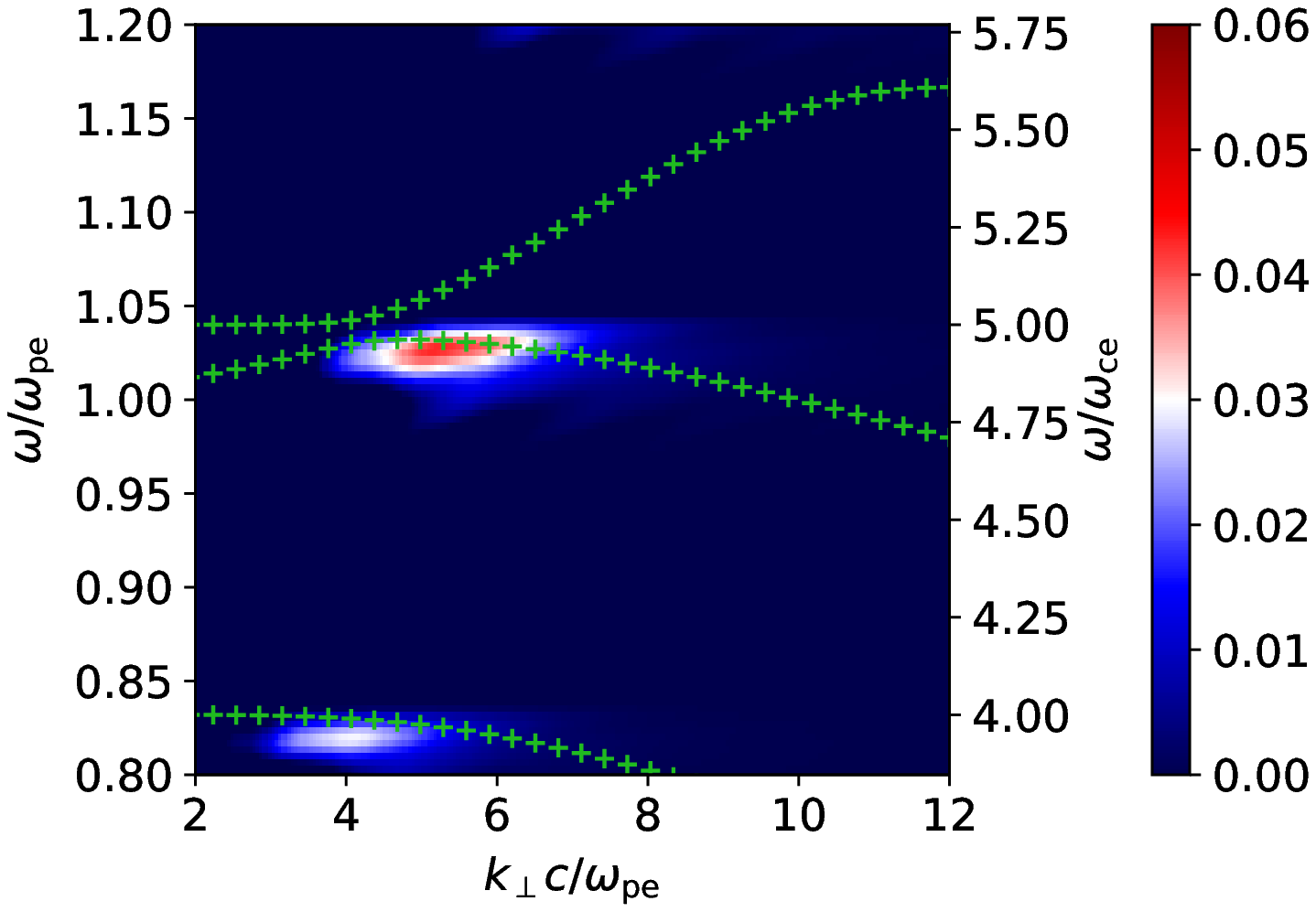}{0.32\textwidth}{(c)}
    }
    \caption{Analytical growth rates normalized to $\omega_\mathrm{ce}$ (blue-red scale) and dispersion branches in dependence
    on the background thermal velocity: a) $v_\mathrm{tb} = 0.007$~$c$, b) $v_\mathrm{tb} = 0.018$~$c$, and c) $v_\mathrm{tb} = 0.030$~$c$.
    The thermal velocity of the hot component $v_\mathrm{t}=0.2$~$c$ and
    $\omega_\mathrm{pe}/\omega_\mathrm{ce}=4.8$ are kept constant in all these cases.} \label{fig4}
\end{figure*}

We also studied the distribution of the growth rates $\gamma$ and dispersion
branches in dependence on the background plasma temperature
(Figure~\ref{fig4}). With increasing the background plasma temperature the
center of the region with the high growth rates remains at the same position,
but the size of this region decreases. Moreover, the point, where dispersion
branches meet and cross the region with the high growth rates, shifts to lower
$k_\perp$. The reason is that for the constant $\lambda = k_\perp^2
v_\mathrm{tb}^2 / \omega_\mathrm{ce}^2$ the component of the wave vector
$k_\perp$ has to decrease when $v_\mathrm{tb}$ increases. It implies that for
the dispersion branches close to the plasma frequency and $v_\mathrm{tb} <
0.018$~$c$ the electrostatic waves are generated at the normal part of the
dispersion branch, while for higher thermal velocities in its anomalous part.

\begin{figure*}[htb!]
    \centering
    \gridline{
            \fig{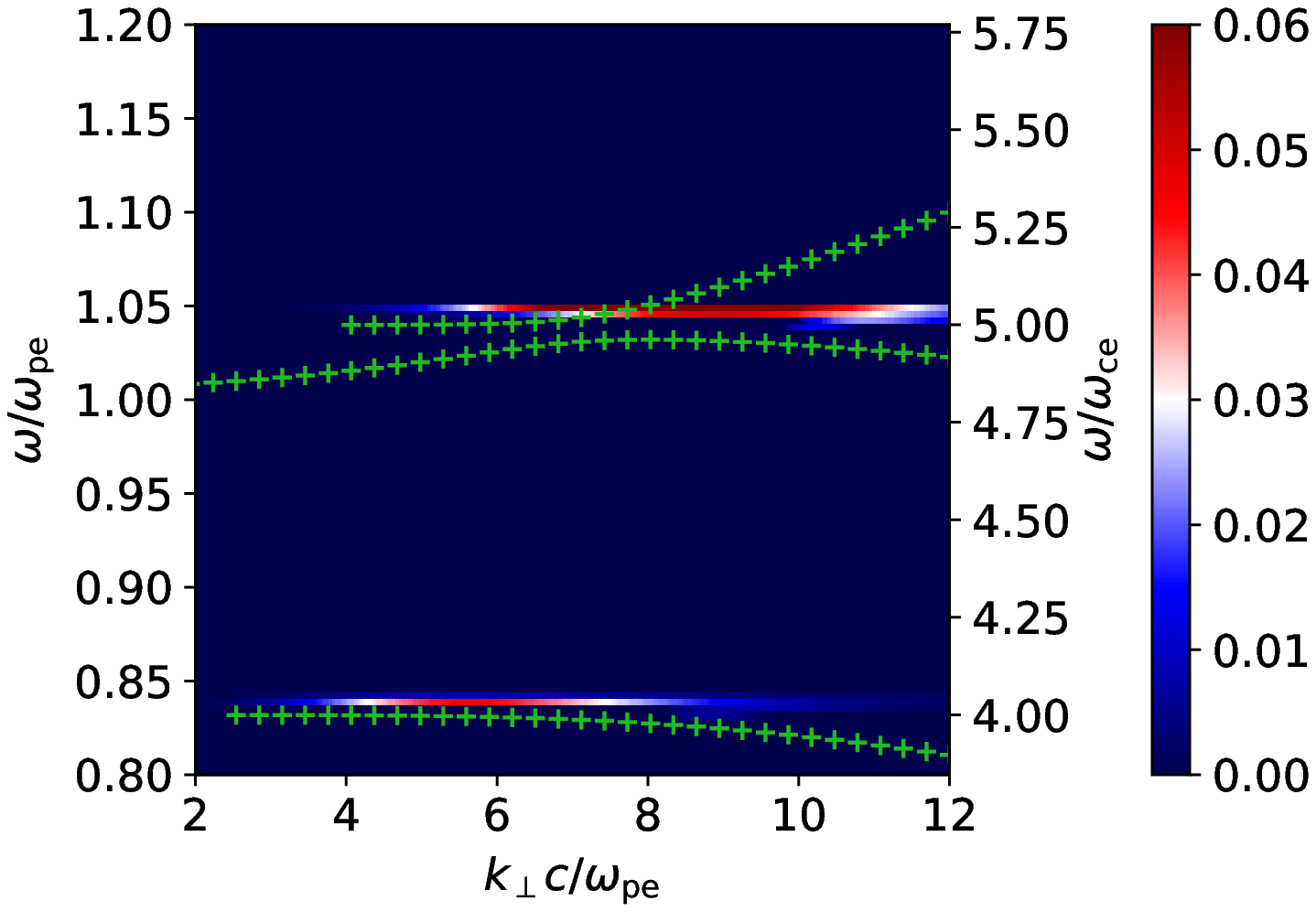}{0.32\textwidth}{(a)}
            \fig{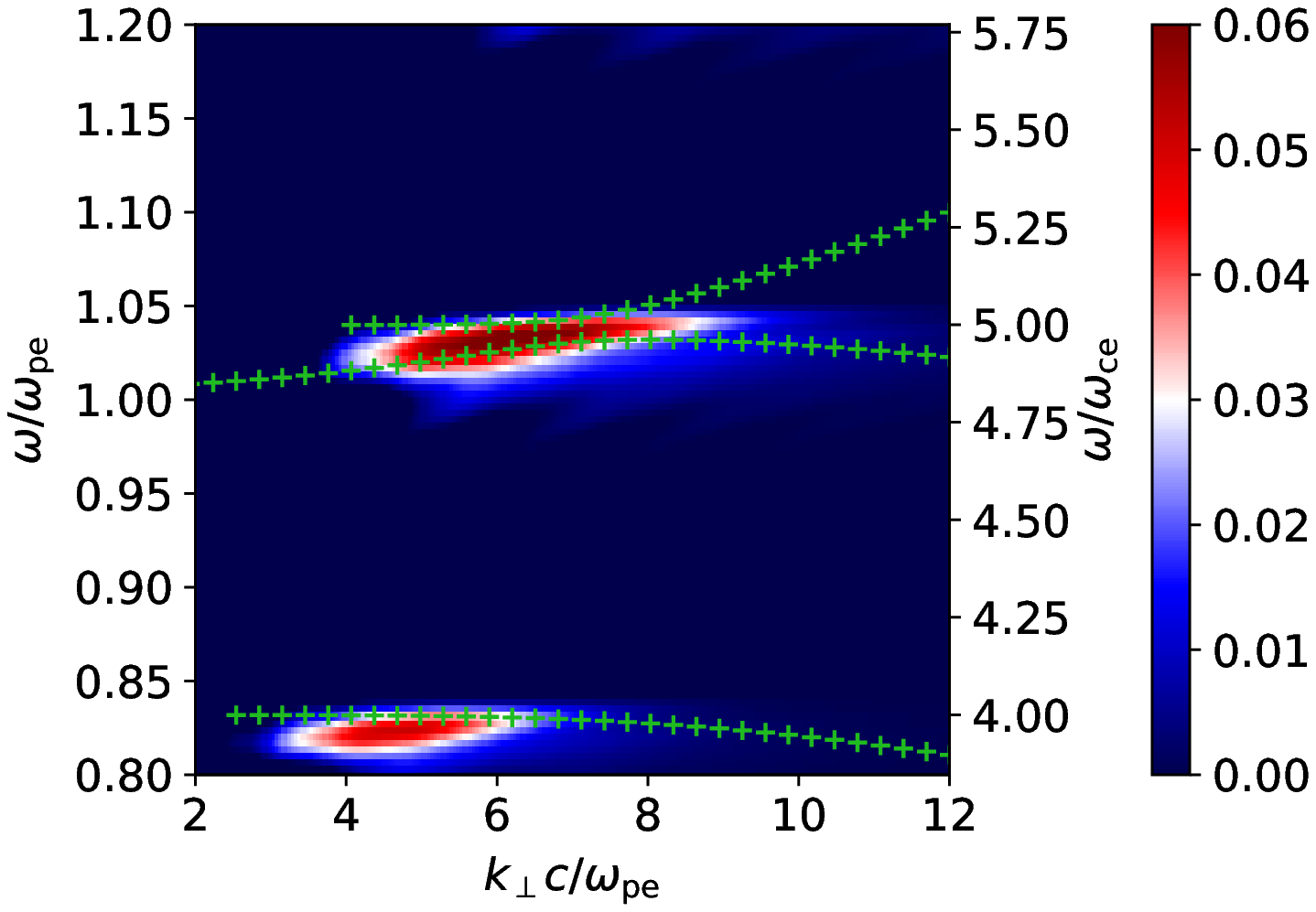}{0.32\textwidth}{(b)}
            \fig{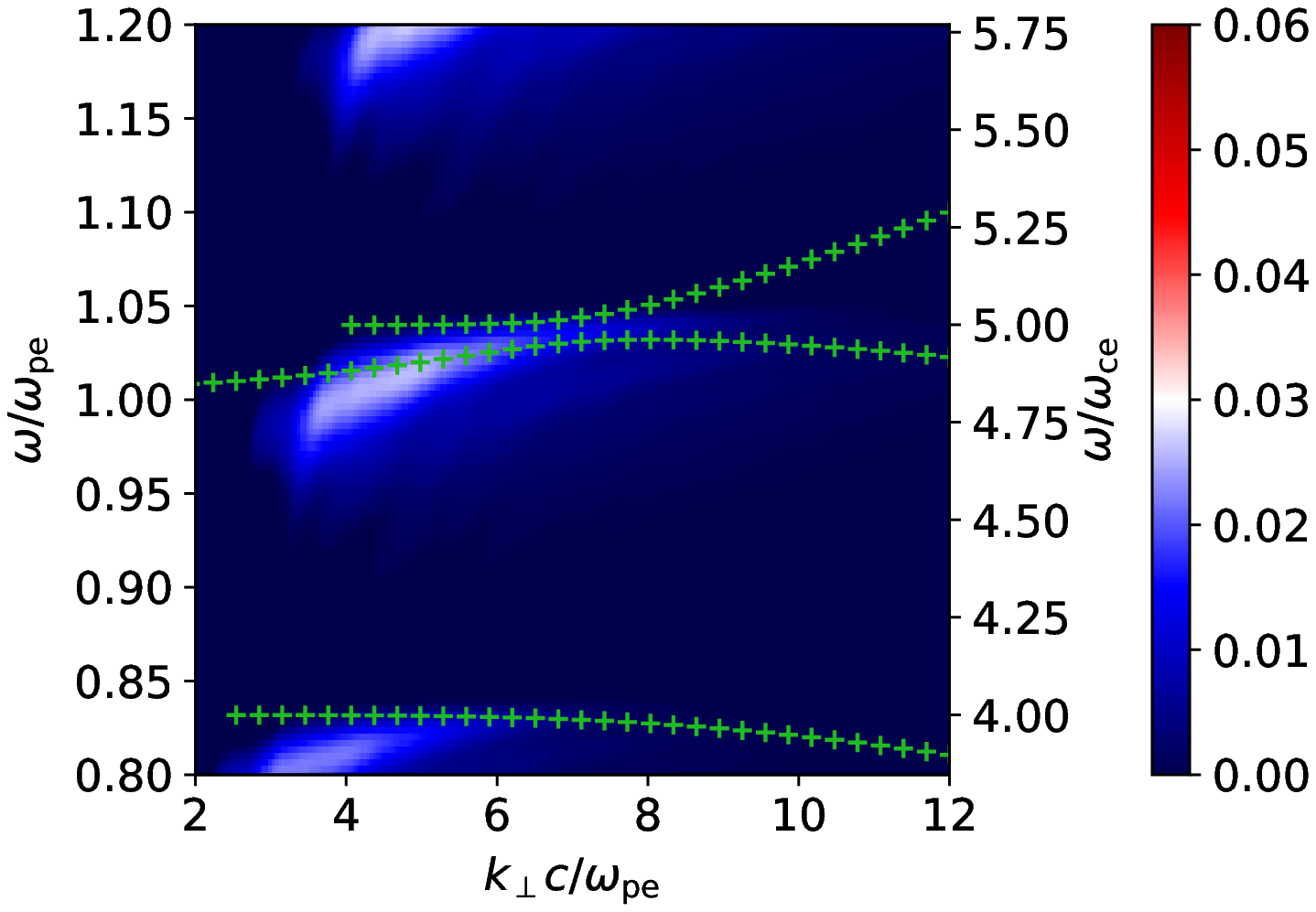}{0.32\textwidth}{(c)}
    }
    \caption{Analytical growth rates normalized to $\omega_\mathrm{ce}$ and dispersion branches in dependence
    on the hot electrons velocity: a) $v_\mathrm{t}=0.1$~$c$, b) $v_\mathrm{t}=0.2$~$c$, and $v_\mathrm{t}=0.3$~$c$.
    $v_{tb}=0.018$~$c$ and $\omega_\mathrm{pe}/\omega_\mathrm{ce}=4.8$
    are kept constant.} \label{fig5}
\end{figure*}

On the other hand, Figure~\ref{fig5} shows the distribution of the growth rates
and dispersion branches in dependence on the "thermal" velocity of hot
electrons $v_\mathrm{t}$. With increasing this velocity the position of the
region with high growth rates shifts to lower $k_\perp$. For a constant
argument in the Bessel function in Equation~\ref{eq7}, const$ = \Gamma v_\perp
k_\perp / \omega_\mathrm{ce}$, $k_\perp$ decreases as $v_\perp$ increases. The
region with high growth rates expands along the frequency axis with increasing
$v_\mathrm{t}$. However, the maximal values of $\gamma$ in the region center
decrease. The reason is that the value of the term $\partial f(v_\perp,
v_\parallel)/\partial v_\perp$ decreases with increasing of $v_\mathrm{t}$. For
temperatures $v_\mathrm{t} < 0.15$~$c$ the area of the region with high growth
rates $\gamma$ is so narrow that the integrated growth rate $\Gamma$ is without
a distinct peak.

\section{Numerical growth rates in the PIC model}

We make simulations using a 3D Particle-in-Cell (PIC) relativistic model
\citep{1985stan.reptR....B,matsumoto.1993,2008SoPh..247..335K,2018A&A...611A..60B}
with multi-core Message Passing Interface (MPI) parallelization. Further
details can be found in \citet[p.67-84]{matsumoto.1993} and on the link
below.\footnote{https://www.terrapub.co.jp/e-library/cspp/text/10.txt.}

The model size is $48\Delta\times48\Delta\times16\Delta$ in $x,y,z$-directions
respectively. The generated electrostatic waves are in $x-y$ plane, $z$
coordinate corresponds to the magnetic field direction. One run takes 80~000
time steps with the time step $\omega_\mathrm{pe} t = 0.025$. The electron
distribution function is DGH distribution, the number of electrons per cell is
$n_\mathrm{e} = 960$, the ratio of densities of the background plasma and hot
electrons is $n_\mathrm{e}/n_\mathrm{h}$ = 32,
$\omega_\mathrm{pe}/\omega_\mathrm{ce}$ = 4-5.3. Other parameters are as in
Table~\ref{tab1}.

\begin{figure}[htb!]
\centering
\includegraphics[width=0.45\textwidth]{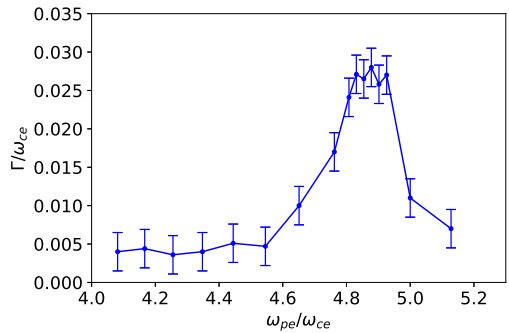}
\caption{Growth rates from PIC simulations as a function of the ratio $\omega_\mathrm{pe}/\omega_\mathrm{ce}$
for $v_\mathrm{t} = 0.2~c, v_\mathrm{tb} = 0.018~c$. Compare it with $\Gamma$ in Figure~\ref{fig1}.}
\label{fig6}
\end{figure}

Varying the ratio of $\omega_\mathrm{pe}/\omega_\mathrm{ce}$ in the
range 4.0-5.3 we made the PIC simulations and estimated the growth rates from the
growth of the electrostatic wave energy (Figure~\ref{fig6}). Note that these
growth rates correspond to the integrated growth rates $\Gamma$ in the
analytical approach. The profile of the growth rate and the growth rate values
are very similar to that presented in Figure~\ref{fig1}. In the range of
$\omega_\mathrm{pe}/\omega_\mathrm{ce}=4.1-4.5$ the growth rate is very weak,
and at $\omega_\mathrm{pe}/\omega_\mathrm{ce}=4.88$ there is the growth rate
maximum. It is at slightly higher values of
$\omega_\mathrm{pe}/\omega_\mathrm{ce}$ comparing with the maximum $\Gamma$
shown in Figure~\ref{fig1}.

\section{Discussion and conclusions}

In this paper we computed two types of the growth rate: a) the growth rate
$\gamma$ that corresponds to one point in the $\omega - k_\perp$ domain, i.e.,
located at one specific dispersion branch and b) the growth rate $\Gamma$
integrated over the upper-hybrid band, where we took into account the fact that
in real conditions as well as in PIC simulations the instability can start
simultaneously not only at one point in the $\omega - k_\perp$ domain, but in
some area in this domain and even on several dispersion branches. While the
maximal growth rate $\gamma/\omega_{ce}$ is found as about 0.06, the
maximal integrated growth rate $\Gamma/\omega_{ce}$ is about 0.03. The
maximal growth rate $\gamma$ is similar to those presented by
\citet{1975SoPh...43..431Z,1986ApJ...307..808W,2017A&A...555A...1B}, but due to
various thermal velocities, plasma densities and magnetic fields used in these
papers this comparison is not straightforward. For example, for the parameters
as in our Table~\ref{tab1}, \citet{1975SoPh...43..431Z} and also
\citet{2009R&QE...52...88Z} found the maximal growth rate as
$\gamma/\omega_\mathrm{ce} \approx 0.03$, which is twice smaller
than our maximal growth rate $\gamma/\omega_\mathrm{ce} \approx 0.06$. The
difference in these values is probably due to different computational
precisions in these studies.

The integrated growth rate is the parameter allowing a direct
comparison with the PIC simulations, contrary to the growth rates at one branch
of the electrostatic waves
\citep{1975SoPh...43..431Z,1986ApJ...307..808W,2005A&A...438..341K,2009R&QE...52...88Z,2017A&A...555A...1B},
see also \cite{1990A&AS...85.1141A}. In computations of the integrated growth
rate we use the "characteristic width" of dispersion branches. We
think that this "characteristic width" is a result of thermal
fluctuations of the electron plasma density. We found that the
"characteristic width" can differ for different dispersion branches.
Just these differences lead to the integrated growth rate $\Gamma$ which agrees
to that computed by PIC simulations.  When we would use the constant
"characteristic width" of dispersion branches ($\sigma = 1$) then the
integrated growth rate would be without peaks at resonances.

We found that the profile of the integrated growth rate $\Gamma$ obtained by
analytical calculations and that in PIC simulations are very similar and their
maxima are at almost the same value of $\omega_\mathrm{pe}/\omega_\mathrm{ce}$.
This difference can be explained by slightly different positions of dispersion
branches in the analytical and numerical approaches. Namely, the condition
$n_\mathrm{h} \ll n_\mathrm{e}$ which is used in the analytical approach, in
PIC simulations is difficult to fulfill, which has an impact on positions of
dispersion branches in PIC model. We found that the growth rate peak in PIC
simulations is broader than that in the analytical approach. We think that it
is because in the PIC model there can also be the electrostatic waves with
$k_\parallel \neq 0$.

We also compared the maximal $\Gamma$ from the present PIC simulations with
that in our previous paper \citep{2018A&A...611A..60B}. We found that due to an
error in the growth-rate normalization in our previous paper the growth rate in
this paper was overestimated 20 times. Considering this correction the maximal
growth rate $\Gamma$ from the present paper agrees to that in the paper by
\citet{2018A&A...611A..60B}.

In dispersion diagrams with dispersion branches, calculated analytically, we
showed how the plasma parameters influence their positions in the $\omega -
k_\perp$ domain. We found that when the dispersion branch with the sufficient
"characteristic width" and length crosses the region with high growth
rates then the integrated growth rate $\Gamma$ is high.

Varying the plasma parameters, we showed that in some range of
$\omega_\mathrm{pe}/\omega_\mathrm{ce}$ the dispersion branches can meet and
change the form to that with a knee. We found that sometimes a dominant
contribution to the integrated growth rate $\Gamma$ comes from the normal part
of the dispersion branch and sometimes from the anomalous part. Moreover, when
some branches meet then it is sometimes difficult to distinguish a type of the
electrostatic wave.

We found that in the upper-hybrid band there can be several dispersion branches
of the electrostatic waves perpendicular to the magnetic field with normal and
anomalous dispersions. We showed that sometimes two branches can even meet in
the $\omega-k_{\perp}$ domain. In the double plasma resonance (DPR) models of
radio zebras only the instability of the upper-hybrid waves with the normal
dispersion (Eq.~\ref{eq0}) is considered. Thus, in new zebra models,
instabilities on all branches in the upper-hybrid band should be taken into
account. Although  it complicates the models, especially analytical ones, they
will be more realistic than the present models. New zebra models need to
describe the processes in the whole upper-hybrid band, because for the zebra
emission the whole electrostatic wave energy, generated on different dispersion
branches, is important. PIC models naturally solve this problem, but they also
need to be improved.

\acknowledgements We thank an anonymous referee for valuable comments.
We acknowledge support from Grants 17-16447S, 18-09072S and 19-09489S of the
Grant Agency of the Czech Republic. This work was supported by The Ministry of
Education, Youth and Sports from the Large Infrastructures for Research,
Experimental Development and Innovations project "IT4Innovations National
Supercomputing Center -- "LM2015070".

\software{\href{https://python.org}{Python}, \href{https://scipy.org}{SciPy}, \href{https://jupyter.org}{Jupyter}}

\end{document}